\documentclass[preprint,12pt]{elsarticle}
\usepackage{amssymb}
\usepackage{mathrsfs}
\usepackage{amsmath}

\usepackage{amssymb}

\journal{Annals of Physics}

\begin{document}

\begin{frontmatter}



\title{Integrable wave function, describing space-time evolution of alpha-decay}


\author{A.Ya.Dzyublik}

\address{Institute for Nuclear Research, National Academy of Sciences of Ukraine,\\
avenue Nauki, 47, Kyiv 03680, Ukraine}

\begin{abstract}
In the framework of decay theory of Goldberger and Watson we
treat $\alpha$-decay of nuclei as a transition caused by a residual interaction between the initial
unperturbed bound state and the scattering states with  $\alpha$-particle.
The integrable wave function for the $\alpha$-decay is derived. The $\alpha$-particle is described by the wave packet, having
 small amplitude inside the nucleus and exponentially growing in external region up to  the $\alpha$-wave
front.   The Moshinsky's distortions of the $\alpha$-wave front
are analyzed.
  It is found  that the energy of the decaying level does not satisfy commonly accepted Bohr-Sommerfeld
   quantization rule for the quasibound levels.
  Only  far from this condition the decay rate turns out to be determined by the Gamov's factor for the  barrier penetrability.
 The derived general   expression for the decay rate is approximated by the familiar quasiclassical formula.

\end{abstract}


\end{frontmatter}


\section{Introduction}
For long time the resonant scattering and decays are intensively discussed in the literature. As
far as in 1928 Gamov \cite{Gamov} and later Condon and Gurney \cite{Condon} explained the experimental data on $\alpha$-decay by tunneling of  $\alpha$-particles
through the Coulomb barrier.  The decaying state  of the parent nucleus is characterized by very small width $\Gamma$ compared to the energy  $Q$ released during
the $\alpha$-decay.
Due to this fact   Gamov \cite{Gamov} reduced the time-dependent task of decay to the stationary Schr\"{o}dinger
equation for the discrete  state of the $\alpha$-particle, described by  the wave function $\psi(r)$ with the energy $E_0$.  The necessity to provide decay of the
nucleus forced Gamov and his followers to replace $E_0$ by the complex energy $E_0-i\Gamma/2$ with $\Gamma$ standing for the width of the level. The corresponding
wave number of the emitted $\alpha$-particle $\kappa\sim \sqrt{E}$ is also complex: $\kappa=\kappa'+i\kappa''$, where $\kappa'>0, \;\kappa''< 0$.
As a consequence, the  wave function, which describes relative motion of nuclear fragments,
diverges with growing  distance $r$ between them,  becoming  not square-integrable, i.e. its norm $||\psi(r)||=\infty$. This contradicts to the probabilistic
interpretation of the wave function. Moreover, such functions are not vectors of the traditional Hilbert space of square-integrable functions with the Hermitian
operators operating there  \cite{Neiman}. Instead, these functions are eigenfunctions of the non-Hermitian Hamiltonians having complex eigenvalues $E$. Example of such
non-Hermitian Hamiltonian, responsible for the cluster decay of nuclei,  has been derived by Silisteanu et al. \cite{Romania}.

Furthermore, it is always demanded that the energy levels of such quasibound states are determined by the quasiclassical Bohr-Sommerfeld quantization rule \cite{Davydov}.
In particular, trying to ensure this restriction  for every direction inside the deformed nuclei,  Ismail et al. \cite{Ismail} imposed  strange requirement that
the depth of the nuclear potential well depends on the  direction of the $\alpha$-particle emission. It is worth to remind that  the Bohr-Sommerfeld condition
indicates position of the  quasistationary levels in the case  of scattering of point-like particles in the potential field (see, e.g., Ref.~\cite{Sitenko}).
These levels manifest themselves as resonances in the scattering cross section.
 The features of such a resonant scattering of structureless $\alpha$-particles in the nuclear potential field have been analyzed by Karpeshin et al.
 \cite{Fedor}.

For validation of the Gamov states in quantum theory  it was introduced a cumbersome formalism of so-called rigged Hilbert space (RHS) [9-14], which already
 adopts the non-Hermitian Hamiltonians and nonintegrable functions.
 In this way the functions $\psi(r)$ for the quasistationary states are interpreted as vectors of RHS, representing also resonances in the scattering.

Sitenko \cite{Sitenko} indicated that  divergence of the $\alpha$-wave function  can be overcome if the particle is described by the wave packet  $\Psi(r,t)$,
formed by the scattering wave functions $\psi_E(r,t)$ with real energies $E$  in the vicinity
of the quasistationary  energy level $E_0,$ being determined by the Bohr-Sommerfeld  rule. He assumed the wave packet to   be localized at $t=0$ inside the nucleus in the
region $0<r<R$. At $t>0$ it begins to spread, leaking from the nucleus. The obtained wave function  $\Psi(r,t)$ again  exponentially grows with $r$, but has a sharp
wave front at the point $r_f=vt$, where $v$ is the velocity of the $\alpha$-particle. This ensures normalization of  $\Psi(r,t)$  to unity.

In more detail leaking of the wave packet, initially localized in the region $0<r<R$, through the potential barrier $V(r)\sim \delta(r-R)$ has been studied in Refs. [15-17] by applying the Moshinsky  function. This function was introduced \cite{Mosh,Mosh1} in the task with a shuttle instantaneously disappearing at $t=0$, that allows spreading of the wave packet in the outer region.

Similar two-potential  model of  decay has been constructed by Gurvitz and Kalberman \cite{Gurvitz}.
 They believed  that at $t<0$ there was a bound level with the energy $E_0$ in a spherically symmetric potential $U(r)$  consisted of
   the potential well at $0\leq r < R$ and the impenetrable  barrier of the constant height $U(r)=V_0>E_0$ at $R\leq r< \infty$. Then at $t=0$
   another potential
$W(r)$   is abruptly switched on, which transforms the initial potential $U(r)$ to more realistic $V(r)=U(r)+W(r)$,  including already the barrier of finite
width.
The authors treated the $W(r)$ as a perturbation, which ensures  decay of the initially bound state. The continuous spectrum in Refs.~\cite{Gurvitz} begins at the brim of the barrier $V_0$
and therefore does not overlap with the bound level $E_0$.  But in this case  the energy conservation law forbids decay of such a bound state. In order to
overcome this difficulty and remove some singularities the authors introduced ad hoc one more potential $\tilde{W}(r)=W(r)+V_0$.

All these models very schematically reproduce physics of the decay.
In reality initially there is no ready $\alpha$-particle. At the initial moment $t=0$ the wave function  $\Psi(r,0)$ is described by the shell
model, which treats all the nucleons as free particles contained in a potential well. The wave functions of such bound states  $\varphi_a$  are eigenfunctions of the unperturbed Hamiltonian $H_0$. Their coupling to the states of the continuum spectrum  $\varphi^+_b$  with ready $\alpha$-particle is realized by means of the residual interaction $V'$, being the difference of the complete Hamiltonian $H$ and $H_0$. At the transition  from  $\varphi_a$ to  $\varphi^+_b$  some potential
energy of the  intrinsic nuclear  motion transforms into  the kinetic energy of  relative motion $E$ of emitted nuclear fragments.
 All this enforces us to conclude that the adequate derivation of the integrable wave function for cluster decay  remains so far a challenge.

In this paper following Goldberger and Watson \cite{Goldberger} we shall split the  Hamiltonian $H$  into the unperturbed Hamiltonian $H_0$
  and the perturbation  $V'$,
\begin{equation}\label{eq:H}
    H=H_0+V'.
\end{equation}
The operator $H_0$ has the eigenfunctions $\varphi_a$ for the
bound states   as well as $\varphi^+_b$ for the continuous
spectrum. Let in the initial moment $t=0$  the parent nucleus be
described by the wave function $\Psi(0)=\varphi_a$, without any
$\alpha$ particle.
 Afterwards the wave function   $\Psi(t)$ attributes components $\sim \varphi^+_b$, whereas the amplitude of
 $\varphi_a$ exponentially attenuates.

In the inverse process of the $\alpha$-scattering by the  nucleus such a state  $\varphi_a$ manifests itself as a resonant compound state, where the energy of the captured $\alpha$-particle is shared among all the nucleons and the particle itself is dissolved in the nucleus.

 \section{Main definitions}
For division of the nuclear Hamiltonian $H$ into the unperturbed
Hamiltonian $H_0$ and perturbation $V'$, responsible for the
decay, we apply  the projection-operator formalism of Feshbach
[23-25]. In this way the Hilbert space of all nuclear wave
functions
  is divided into wave  functions of quasibound states and wave functions of the states of the continuous spectrum.
Next, the projection operators ${\cal Q}$
and ${\cal P}$  are introduced, which act, respectively, on the quasibound and  scattering states,
\begin{equation}\label{}
{\cal P}^2={\cal P},\quad {\cal Q}^2={\cal Q},\quad {\cal P}+{\cal Q}=1,\quad {\cal Q}{\cal P}=0.
\end{equation}
Correspondingly, any wave function of the nuclear system can be represented by the expression
\begin{equation}\label{}
\Psi={\cal P}\Psi +{\cal Q}\Psi,
\end{equation}
while the exact Hamiltonian by
\begin{equation}\label{}
H=H_{{\cal Q}{\cal Q}}+H_{{\cal PP}}+H_{{\cal Q}{\cal P}}+H_{{\cal P}{\cal Q}},
\end{equation}
where $H_{{\cal Q}{\cal Q}}={\cal Q}H{\cal Q}$, etc.
In this way the unperturbed Hamiltonian can be defined as
\begin{equation}\label{}
H_0=H_{{\cal Q}{\cal Q}}+H_{{\cal PP}}
\end{equation}
and the perturbation operator as
\begin{equation}\label{}
V'=H_{{\cal QP}}+H_{\cal PQ}.
\end{equation}
The basis functions $\varphi_a$ of the ${\cal Q}$-subspace are determined by the eigenvalue equation
\begin{equation}\label{}
H_{{\cal Q}{\cal Q}}\varphi_a=\varepsilon_a\varphi_a,
\end{equation}
while the vectors $\varphi_b^+$ of the ${\cal P}$-part of the Hilbert space by
 \begin{equation}\label{}
H_{{\cal P}{\cal P}}\varphi_b^+=\varepsilon_b\varphi_b^+.
\end{equation}
All the functions $\varphi_a$ and $\varphi^+_b$ are orthogonal because $H_0$ is a Hermitian operator. They   form a complete set of basis vectors.
Although the explicit form of the operators $H_0$ and $V$ remains obscure, the Feshbach's approach allows us to understand  main features of the $\alpha$-decay.

Let the charge number of the parent nucleus be $Z$ and the mass number $A$.
The nuclear decay is considered in the c.m. frame. Both potential $V_n(\beta;r)$ and $V_C(\beta;r)$, depending on the deformation parameter $\beta$, can be
expanded in the series in $\beta$. Here we consider the zeroth order terms $V_n(r)$ and $V_C(r)$ representing  spherically symmetric potentials. The corrections to these potentials,
dependent on $\beta$, can be accounted in the coupled-channels formalism \cite{Dzyublik1}.

In the $\alpha$-decay channel we have ready $\alpha$-particle and a daughter nucleus,
whose relative motion is determined by the radius-vector ${\bf r}$, while their intrinsic motion by the coordinates $\xi_{\alpha}$ and $\xi_d$, respectively.
The same variables are used in the cluster model (see, e.g., \cite{Clust}) for any separated  group of two protons and two neutrons as well as $A-4$ nucleons of
the parent nucleus. The parent nucleus in the initial state, formed at $t=0$,
 consists of free nucleons, moving in some central potential field.

Let at $t=0$ this initial state be described by the function $\varphi_a=g_p(\xi, {\bf r})$, where $\xi=\{\xi_{\alpha},\xi_d\}$, while the corresponding eigenvalue
of $H_0$ be  $\varepsilon_a={\cal M}_{p}c^2+{\cal E}_p$, where ${\cal E}_p$ is the  energy of the excited nuclear level,  the subscript  $p$ specifies the spin
$I_p$, its projection $M_p$ on the quantization axis and any other quantum numbers. Hereafter ${\cal M}_{p(d)}$ and ${\cal M}_{\alpha}$ are the masses,
respectively, of the parent (daughter) nucleus and the $\alpha$-particle being in the ground state.

In the $\alpha$-decay channel the unperturbed Hamiltonian $H_0$ is a sum of the kinetic energy operator of the relative motion of the fragments $K$, their
potential energy $V(r)$, Hamiltonians for internal motion of the daughter nucleus $H^{(d)}_{{\textrm{\scriptsize {in}}}}$ and the $\alpha$-particle
$H^{(\alpha)}_{{\textrm{\scriptsize {in}}}}$:
\begin{eqnarray}\label{eq:2h}
H_0=K+V(r)+H^{(\alpha)}_{{\textrm{\scriptsize {in}}}}
+H^{(d)}_{{\textrm{\scriptsize {in}}}}, \qquad K=-\frac{\hbar^2}{2\mu}\Delta_{\bf r},
\end{eqnarray}
where the reduced mass $\mu={\cal M}_{d}{\cal M}_{\alpha}/({\cal M}_d+{\cal M}_{\alpha})$.

The eigenfunctions of $H_0$ are
 \begin{equation}\label{eq:t2}
 \varphi_{b}^+(\xi,{\bf r})=\psi^+_{{\boldsymbol \kappa}}({\bf r})g_{d}(\xi_d)g_{\alpha}(\xi_{\alpha}).
\end{equation}
These wave functions must be yet antisymmetryzed \cite{R2}.
The factors $g_{d}(\xi_d)$ and $g_{\alpha}(\xi_{\alpha})$ describe internal motion of the clusters and are determined by the equations
\begin{eqnarray}\label{eq:3f}
 H^{(\alpha)}_{{\textrm{\scriptsize {in}}}}g_{\alpha}(\xi_{\alpha})={\cal M}_{\alpha}c^2g_{\alpha}(\xi_{\alpha}),  \\
  H^{(d)}_{{\textrm{\scriptsize {in}}}}g_d(\xi_d)=({\cal M}_dc^2+{\cal E}_d)g_d(\xi_d), \nonumber
 \end{eqnarray}
where ${\cal E}_{d}$ is the excitation
 energy of the daughter nucleus, the subscript $d$ of $g_d(\xi_d)$ includes
 spin $I_d$, its projection $M_d$, etc.

The function $\psi^+_{{\boldsymbol \kappa}}({\bf r})$, responsible for the relative motion, satisfies the Shr\"{o}dinger equation
  \begin{equation}\label{eq:Schr}
 \left[\frac{\hbar^2}{2\mu}\Delta_{\bf r}-V(r)+E \right]\psi^+_{{\boldsymbol \kappa}}({\bf  r})=0,
\end{equation}
where $E=\hbar^2\kappa^2/2\mu$ is the energy of the relative motion of fragments.

The unperturbed energies associated with $ \varphi_{b}^+(\xi,{\bf r})$    are
\begin{equation}\label{eq:112}
\varepsilon_b=({\cal M}_d+{\cal M}_{\alpha})c^2 +{\cal E}_{d}+E.
\end{equation}

The  Coulomb field for bare uniformly charged nuclei  is given by
\begin{equation}\label{eq:Co1}
V_{{\textrm{\scriptsize{C}}}}( r)_{{\textrm{\scriptsize {b}}}}= \left\{\begin{array}{lll} \frac{(Z-2)e^2}{R}
 \left[3-\frac{r^2}{R^2}\right],&\;&\;0\leq r<R,\\
\frac{2(Z-2)e^2}{r} ,&\;&\;r>R,  \end{array}\right.,
\end{equation}
where $R$ is the nuclear radius.

Further we shall only consider the decay of nuclei, surrounded by electrons.
In this case   $\alpha$-particle moves  in the field
 \begin{eqnarray}\label{eq:2}
V(r)= V_n(r)+  V_{{\textrm{\scriptsize {C}}}}(r),
\end{eqnarray}
 where  $V_n(r)$ stands for the nuclear potential well and $ V_{{\textrm{\scriptsize {C}}}}(r)$ for the effective Coulomb field. At small distances, when the
 $\alpha$-particle moves inside the nucleus or under the  barrier, the Coulomb contribution up to small correction $\sim r^2$  is \cite{Zinner,Dzyublik}
\begin{equation}\label{}
V_{{\textrm{\scriptsize{C}}}}( r)\approx V_{{\textrm{\scriptsize{C}}}}( r)_{{\textrm{\scriptsize {b}}}}-\Delta Q,
\end{equation}
where   $\Delta Q$ is the energy transferred to electrons. In nonmetallic targets  $\Delta Q= B_p-B_d$, where   $B_p$ and $B_d$ are   the electron binding
energies of the parent and daughter atoms. The conductivity electrons give small correction  \cite{Dzyublik}.

Respectively, the nuclear energies are related by $\varepsilon_b \approx \varepsilon_a-\Delta Q$ with uncertainty  of the order of the decay width $\Gamma$.
The relative energy $E$ of  clusters in the $d$th channel is spread about the mean energy $E_d$. For the decay by screened nuclei \cite{Zinner,Dzyublik}
 \begin{equation}\label{eq:e}
   E_d= Q_{d}-\Delta Q,
  \end{equation}
where
\begin{equation}\label{eq:qq}
 Q_d=({\cal M}_p-{\cal M}_{\alpha}-{\cal M}_d)c^2+{\cal E}_p-{\cal E}_d,
\end{equation}
is the average nuclear  energy released in this decay.

\section{Scattering wave functions}
 The functions $\psi^+_{{\boldsymbol \kappa}}({\bf r})$  are normalized by
\begin{equation}
 \langle \psi^+_{{\boldsymbol \kappa}'}({\bf r})|\psi^+_{{\boldsymbol \kappa}}({\bf r})\rangle =\delta({\boldsymbol \kappa}'- {\boldsymbol \kappa}).
\end{equation}
In the asymptotic region, $r \to \infty$, they are represented by  a sum
 of the incident wave $(2\pi)^{-3/2}e^{i{\boldsymbol \kappa}{\bf r}}$ and a spherical outgoing wave $\sim\frac{1}{r}e^{i\kappa r}$.

The $\psi^+_{{\boldsymbol \kappa}}({\bf r})$ can be expanded in  partial waves \cite{Goldberger}:
\begin{equation}\label{eq:t3}
\psi^+_{{\boldsymbol \kappa}}({\bf r})=
\sum_{l=0}^{\infty}\sum_{m=-l}^{l}i^l
e^{i\delta_l(\kappa)}\frac{w_l(\kappa;r)}{\kappa r}
Y_{lm}^{*}(\hat{{\boldsymbol \kappa}})Y_{lm}(\hat{\bf r}),
\end{equation}
where $\hat{{\boldsymbol \kappa}}$  and $\hat{\bf r}$ denote the spherical angles of the vectors ${\boldsymbol \kappa}$ and ${\bf r}$, respectively.
Here the radial functions $w_l(\kappa;r)$ satisfy the equation
\begin{equation}\label{eq:t4}
w''_l(\kappa;r)-\left[ l(l+1)/r^2+v(r)-\kappa^2 \right]w_l(\kappa;r)=0,
\end{equation}
where  the reduced potential
\begin{equation}\label{}
v(r)=2\mu V(r)/\hbar^2.
\end{equation}

The regular functions at $r \to 0$ behave as
\begin{equation}\label{eq:w1}
w_l(\kappa;r)\sim (\kappa r)^{l+1}.
\end{equation}
   With growing $r$ the screened Coulomb potential $V_{{\textrm{\scriptsize {C}}}}(r)$ attenuates faster than a pure Coulomb one. Respectively, at $r \to \infty$
   the functions $w_l(\kappa;r)$ have more simple  asymptotic than the Coulomb functions  \cite{Goldberger}:
\begin{equation}\label{eq:t6}
 w_l(\kappa;r)\approx \sqrt{\frac{2}{\pi}}
\sin\left(\kappa r-\frac{l\pi}{2}+\delta_l(\kappa)   \right),
\end{equation}
where $\delta_l(\kappa) $ stands for the phase shift.
These functions are normalized as follows:
\begin{equation}\label{eq:t7}
\int_{0}^{\infty} w_l(\kappa;r)w_l(\kappa';r)dr=\delta(\kappa-\kappa').
\end{equation}

Since the Hamiltonian of the closed nuclear system $H$ is invariant with respect to rotations it is more appropriate to expand the basis functions
$\varphi^+_b(\xi,{\bf r})$
in terms of the eigenfunctions of the operators ${\bf I}^2$ and $I_z$, where  ${\bf I}={{\bf I}}_d+{\bf l}$ is the total angular momentum operator of the nuclear
clusters and $I_z$ its projection on the quantization axis $z$. Due to such symmetry of $H$ the interaction $V$ couples the states with the same total spin and
its projection, i.e., $I=I_p$ and $M=M_p$.

Spin of the daughter nucleus and the orbital angular momentum ${\bf l}$ are coupled, giving
the eigenfunctions of  ${\bf I}^2$ and $ I_z$:
\begin{equation}\label{eq:t8}
\mathscr{Y}_{IlI_d}^{M}(\xi,\hat{\bf r})=\sum_{mM_d}(lI_d mM_d |IM)Y_{lm}(\hat{\bf r})g_{I_dM_d}(\xi_d)g_{\alpha}(\xi_{\alpha}),
\end{equation}
where $(j_1j_2m_1m_2|jm)$ are the Clebsh-Gordan coefficients.
The reverse transformation is
\begin{equation}\label{eq:t800}
 Y_{lm}(\hat{\bf r}) g_{I_dM_d}(\xi_d)g_{\alpha}(\xi_{\alpha}) =\sum_{IM}(lI_d mM_d |IM)\mathscr{Y}_{IlI_d}^{M}(\xi;\hat{\bf r}).
\end{equation}
Note that the functions $\mathscr{Y}_{IlI_d}^{M}(\xi;\hat{\bf r})$ are analog of the generalized spherical harmonics used by Newton \cite{Newton}.
By inserting (\ref{eq:t3}) into (\ref{eq:t2}) and using  (\ref{eq:t800})
we rewrite the wave function $\varphi^+_b$ as
\begin{eqnarray}\label{eq:t801}
  \varphi^+_b(\xi,{\bf r})=\sum_{IM}\sum_{l=0}^{\infty}\frac{w_l(\kappa;r)}{\kappa r}
    \mathscr{Y}_{IlI_d}^{M}(\xi,\hat{\bf r})
\mathfrak{Y}_{I}^{M^*}(lI_dM_d;\hat{\boldsymbol \kappa}),
\end{eqnarray}
where we introduced the notation
\begin{equation}\label{eq:y}
\mathfrak{Y}_{I}^{M}(lI_dM_d;\hat{\boldsymbol \kappa})=i^{-l}e^{-i\delta_l}\sum_{m=-l}^l (lI_d mM_d|IM)Y_{lm}(\hat{\boldsymbol \kappa}).
\end{equation}

 \section{Quasi-classical  approximation}
Let us solve the radial equation (\ref{eq:t4}) in the
quasi-classical (WKB) approximation. It does not "work" at
$r\approx 0$, when the effective potential quickly changes at the
distance of the order of the wavelength \cite{Davydov}. In order
to overcome this obstacle Langer \cite{Langer}  replaced the
variable $r$ by $x$
\begin{equation}\label{eq:W4}
 r=\kappa^{-1}e^x.
\end{equation}
New coordinate $x$ varies on the whole axis from $-\infty$ to $\infty$ with
$x\to -\infty$ corresponding to the point $r=0$.
By substitution
\begin{equation}\label{eq:W5}
 w_l(\kappa;r)=e^{x/2}y_l(x)
\end{equation}
 Eq.~(\ref{eq:t4})  transforms to
\begin{equation}\label{eq:WKB5}
y''_l(x)+q^2(x)y_l(x)=0,
\end{equation}
with
\begin{equation}\label{eq:WKB6}
q^2(x)=e^{2x}\left( 1-\frac{v(x)}{\kappa^2}\right)-\left(l+\frac{1}{2}\right)^2.
\end{equation}

The Eq.~(\ref{eq:WKB5}) can be solved already in the WKB approximation.
Reverse transformation to coordinate $r$ gives us
\begin{equation}\label{k1}
q^2(x)=r^2k_l^2(r),
\end{equation}
where $k_l(r)$ is the quasi-classical wave number
\begin{equation}\label{eq:k2}
 k_l(r)=\sqrt{\kappa^2- v_{{\textrm{\scriptsize {eff}}}}(r) },
\end{equation}
expressed in terms of the reduced effective potential
\begin{equation}\label{}
v_{{\textrm{\scriptsize {eff}}}}(r)=\frac{2\mu}{\hbar^2}V_{{\textrm{\scriptsize {eff}}}}(r)
\end{equation}
with
\begin{equation}\label{}
V_{{\textrm{\scriptsize {eff}}}}(r)=V(r)+\frac{\hbar^2(l+1/2)^2}{2\mu r^2}.
\end{equation}

 The classical  turning points $x_1$, $x_2$ and $x_3$ on the axis $x$ are  the roots of the equation $q(x)=0$.
They are related, respectively,  to the turning  points $r_1$, $r_2$ and $r_3$, where $k_l(r_i)=0$ (see  Fig.1).
 \begin{figure}[t]
\vspace{0cm}
\centerline{\includegraphics[height= 8cm, width= 8cm]{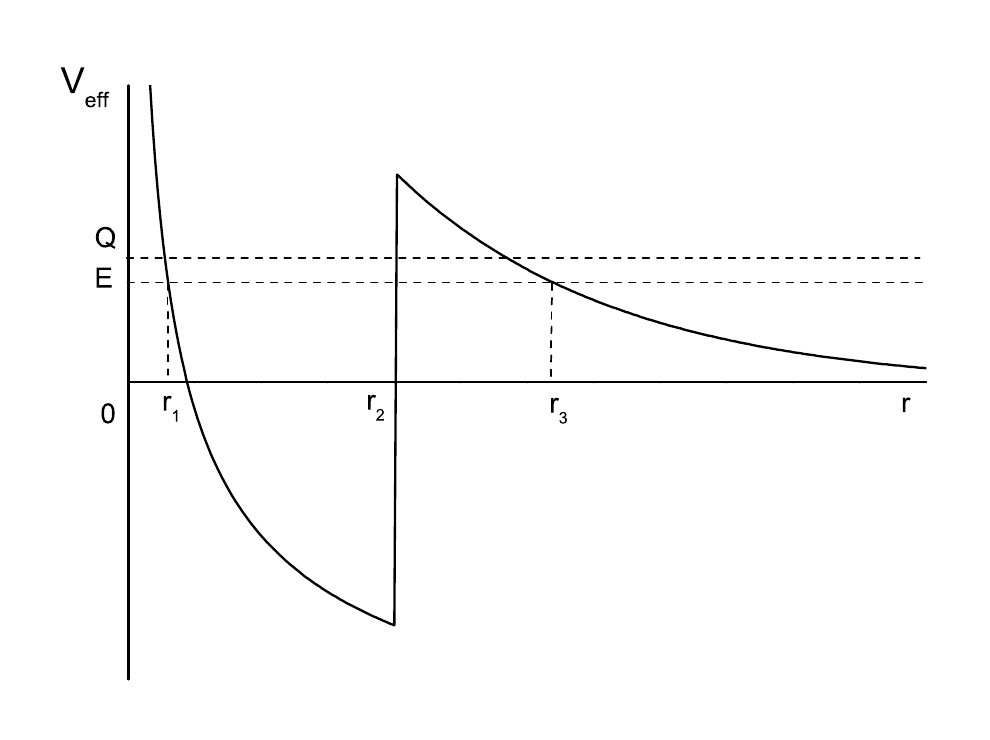}}
\vspace{-1cm}
\caption{\label{fig1} The effective potential energy for relative motion of the $\alpha$-particle and  daughter nucleus, which incorporates the nuclear square
potential well, screened Coulomb interaction and the Lange's centrifugal barrier.
   $E_d=Q_d-\Delta Q$ is the mean kinetic energy of relative motion   at $r \to \infty$ in the $d$th channel, $Q_d$ is the mean nuclear energy released during
   $\alpha$-decay,  $\Delta Q$ is a part
of $Q_d$ absorbed by electrons. The classical turning points are denoted as $r_1,\;r_2,\; r_3$.}
\end{figure}
Under the barrier on the left-hand side  of the turning point $x_1$ the regular WKB solution is represented by the attenuating exponent:
\begin{equation}\label{eq:W7}
y_l(x)=\frac{C_l}{\sqrt{|q(x)|}}\exp\left(-\int_{x}^{x_1}|q(x')|dx' \right), \quad x<x_1.
\end{equation}
Using standard matching rules one finds the function in the potential well, where $x_1<x<x_2$:
\begin{equation}\label{eq:W8}
y_l(x)=\frac{2C_l}{\sqrt{q(x)}}\cos\left(\int_{x_1}^{x}q(x')dx' -\frac{\pi}{4}  \right),
\end{equation}
as well as under the Coulomb barrier, where $x_2<x<x_3$:
\begin{eqnarray}\label{eq:W9}
y_l(x)=\frac{C_l}{\sqrt{|q(x)|}}\left\{\cos\alpha_l
e^{-S_l(Q)}\exp\left(\int_x^{x_3}|q(x')|dx' \right)\right. \nonumber \\
\left.-2\sin\alpha_l e^{S_l(Q)}\exp\left(-\int_x^{x_3}|q(x')|dx' \right)
\right\}.\qquad\qquad
\end{eqnarray}
Here are introduced  the action
\begin{equation}\label{eq:660}
S_l(Q)=\int^{x_3}_{x_2}|q(x')|dx'
\end{equation}
and the angle
\begin{equation}\label{eq:610}
\alpha_l=\int_{x_1}^{x_2}q(x)dx -\frac{\pi}{2}.
\end{equation}

Behind the barrier when $x>x_3$
\begin{eqnarray}\label{eq:w55}
y_l(x)=-\frac{C_l}{\sqrt{q(x)}}
\left\{\cos\alpha e^{-S_l}\sin\left(\int_{x_3}^x q(x')dx'-\frac{\pi}{4}\right)\right.\nonumber \\ \left.
+4\sin\alpha e^{S_l}
\cos\left(\int_{x_3}^xq(x')dx'-\frac{\pi}{4}\right)
\right\}.\qquad\qquad
\end{eqnarray}

Let us return now to the radial coordinate $r$ by means of Eqs.~(\ref{eq:W4}), (\ref{eq:W5}). The action (\ref{eq:660}) and the angle (\ref{eq:610})  are
rewritten then as
\begin{equation}\label{eq:s}
S_l(Q)=\int_{r_2}^{r_3} |k_l(r)|dr
\end{equation}
and
\begin{equation}\label{eq:alpha}
\alpha_l=\int_{r_1}^{r_2}k_l(r)dr-\frac{\pi}{2}
\end{equation}
with $k_l(r)$  defined by Eq.~(\ref{eq:k2}).

The action (\ref{eq:s}) completely defines the barrier penetrability $e^{-2S_l(Q)}$.
It can be rewritten in more familiar form:
\begin{equation}\label{eq:Gf}
e^{-2S_l(Q)}=\exp\left\{-\frac{2}{\hbar}\int_{r_2}^{r_3}\sqrt{2\mu\left(V_{{\textrm{\scriptsize {eff}}}}(r)-E_d \right)}dr\right\}.
\end{equation}

The  condition $\alpha_l=n \pi$, where $n=1,\;2,\;3\ldots$, is equivalent to equalities
 \begin{equation}\label{eq:Som}
\int_{r_1}^{r_2}k_l(r)dr=\left(n+\frac{1}{2}\right)\pi, \qquad n=1,\; 2,...,
\end{equation}
or
\begin{equation}\label{eq:Som2}
\oint p_l(r)dr=2\pi \hbar\left(n+\frac{1}{2}\right),\qquad p_l=\hbar k_l,
\end{equation}
representing the Bohr-Sommerfeld quantization rule (see, e.g., \cite{Davydov}) to  determine the resonant  energies $E_r$ (the quasi-stationary levels).

Starting from the expression (\ref{eq:w55}), one can write the wave function behind  the Coulomb barrier  $(r>r_3)$ in the form
\begin{equation}\label{eq:W10}
 w_l(\kappa;r)=C_l\left(\frac{\kappa}{k_l(r)}\right)^{\frac{1}{2}}
 X_l\sin\left(\int_{r_3}^r k_l(r)dr +\gamma+\frac{\pi}{4} \right),
\end{equation}
where new parameters $X_l$ and $\gamma$ satisfy the equations
\begin{eqnarray}\label{}
X_l\sin\gamma=\cos\alpha_l e^{-S_l},\qquad X_l\cos\gamma=-4\sin\alpha_l e^{S_l},
\end{eqnarray}
whose solutions are
\begin{equation}\label{}
X_l=\left[16\sin^2\alpha_l e^{2S_l}+ \cos^2\alpha_l e^{-2S_l}
  \right]^{1/2}
\end{equation}
and
\begin{eqnarray}\label{eq:gam}
\gamma = -\arctan\left(\frac{e^{-2S_l}}{ 4\tan\alpha_l}\right).
\end{eqnarray}

At large distance $k_l(r)$ approaches $\kappa$.
Therefore in the asymptotic region the integral  in Eq.~(\ref{eq:W10}) can be rewritten as
\begin{equation}\label{}
\int_{r_3}^r k_l(r)dr=\int_{r_3}^{\infty} [k_l(r)-\kappa]dr +\kappa (r-r_3).
\end{equation}
Afterwards equating the function $w_l(\kappa;r)$, where $r \to \infty$,  to its asymptotic form (\ref{eq:t6}), one gets the amplitude
\begin{equation}\label{eq:aa}
C_l=\sqrt{\frac{2}{\pi}}\frac{1}{X_l}
\end{equation}
and the scattering phase shift
\begin{equation}\label{eq:del}
\delta_l=
\bar{\delta_l} +\gamma,
\end{equation}
where $\gamma$ is given by (\ref{eq:gam}) and
 \begin{equation}\label{}
\bar{\delta_l}  =
\int_{r_3}^{\infty}[k_l(r)-\kappa]dr-\kappa r_3+
\left(l+\frac{1}{2}\right)\frac{\pi}{2}.
 \end{equation}

The amplitude squared $C_l^2$ determines both the rate of reactions with  $\alpha$ projectiles and  the $\alpha$-decay rate.
Therefore it is interesting to examine its behavior as a function of the deviation from the resonance $\Delta\alpha_l=\alpha_l-n\pi$.
 By assumption, the tunneling probability through the Coulomb barrier $e^{-2S_l}<<1$. If $|\Delta \alpha_l|>>e^{-2S_l}/4$, the   angle $\gamma$ and the amplitude
 $C_l$  are very small:
\begin{equation}\label{eq:amp}
\gamma=-\frac{1}{4} e^{-2S_l}\cot\alpha_l, \qquad\qquad
C_l=\left(\frac{2}{\pi}\right)^{\frac{1}{2}}\frac{e^{-S_l}}{4\sin\alpha_l}.
\end{equation}

In the opposite case, when $\Delta\alpha_l\approx 0 $, the
$C_l^2$ versus $\Delta\alpha_l$ is described by the Lorentzian function
 \begin{equation}\label{eq:res1}
 C_l^2=C^2_{l,res}\frac{\left(e^{-2S_l}/4\right)^2}{(\Delta\alpha_l)^2+\left(e^{-2S_l}/4\right)^2},
 \end{equation}
 where $C^2_{l,res}=2e^{2S_l}/\pi$ is the resonant value of $C^2_l$.
Let us find now the dependence of $C_l^2$ on energy in the vicinity of the resonance. By using two first terms of the Tailor series
 \begin{equation}\label{}
\alpha_l(E)=\alpha_l(E_r)+ \left(\frac{d\alpha_l}{dE}\right)_{r}(E-E_r)+\ldots,
 \end{equation}
we rewrite  Eq.~(\ref{eq:res1}) as
\begin{equation}\label{eq:cs}
C^2_l(E)=C^2_{l,res}\frac{(\Gamma_r/2)^2}{(E-E_r)^2+(\Gamma_r/2)^2},
\end{equation}
where $\Gamma_r$ is the width of the resonance,
\begin{equation}\label{eq:res}
\Gamma_r =\frac{e^{-2S_l}}{2\left(\frac{d\alpha_l}{dE}\right)_{r} }.
\end{equation}
Thus, the curve (\ref{eq:cs}) has  extremely narrow width $\Gamma_r \sim e^{-2S_l}$.

In the same resonant case $\gamma$ is determined by the formula
\begin{equation}\label{}
\gamma= - \arctan \left(\frac{e^{-2S_l}}{4\Delta\alpha_l} \right).
\end{equation}
Substituting it into Eq.~(\ref{eq:del}) one gets the well-known expression (see, e.g., \cite{Sitenko}) for the phase shift in the case of  isolated resonance:
 \begin{equation}\label{}
\delta_l=\bar{\delta_l}  -\arctan \left(\frac{\Gamma_r/2}{E-E_r} \right).
\end{equation}
From here we see that $\bar{\delta_l}$ means the phase shift far from the resonance.

Similarly, by using (\ref{eq:W4}), (\ref{eq:W7}) and (\ref{eq:W9}), one finds the WKB wave
function inside the nucleus at $r_1<r<r_2$
\begin{equation}\label{eq:W13}
w_l(\kappa;r)=2C_l\sqrt{\frac{\kappa}{k(r)}}
\cos\left(\int_{r_1}^r k_l(r')dr'-\frac{\pi}{4} \right)
\end{equation}
and under the  centrifugal barrier as $0<r<r_1$
\begin{equation}\label{eq:W14}
w_l(\kappa;r)=C_l\sqrt{\frac{\kappa}{k_l(r)}}
\exp\left(-\int_r^{r_1}|k_l(r')|dr'\right).
\end{equation}

  Once the factor $e^{-2S_l}<<1$, the probability that  $\Delta\alpha_l$  lies  in the narrow resonant interval $\sim e^{-2S_l}$   will be extremely low.
  Therefore below   for  the amplitude $C_l$ we use Eq.~(\ref{eq:amp}).
Then   under the Coulomb barrier $w_l(\kappa;r)$  exponentially grows with  $r$ changing from $r_2$ to $r_3$:
\begin{eqnarray}\label{eq:W15}
w_l(\kappa;r)=\left(\frac{2}{\pi}\right)^{\frac{1}{2}}e^{-S_l}
\sqrt{\frac{\kappa}{|k_l(r)|}}
  \exp\left(\int_{r_2}^{r}|k_l(r')|dr'\right).
\end{eqnarray}

Behind the barrier, $r> r_3$, it is given by
\begin{equation}\label{eq:W16}
w_l(\kappa;r)=
\left(\frac{2}{\pi}\right)^{\frac{1}{2}}\sqrt{\frac{\kappa}{k_l(r) }}
\sin \left(\int_{r_3}^r k_l(r')dr' +\frac{\pi}{4}  \right).
\end{equation}

The  irregular WKB solution $\tilde{w}_l(\kappa;r)$ of the Schr\"{o}dinger equation
(\ref{eq:t4}) is calculated in the same manner. It  diverges at $r\to 0$,
\begin{equation}\label{eq:ww}
\tilde{w}_l(\kappa;r) \sim
\exp\left(\int_r^{r_1}|k_l(r')|dr'\right),
\end{equation}
and has the asymptotic
\begin{equation}\label{eq:z6}
\tilde{w}_l(\kappa;r)\sim -\sqrt{\frac{2}{\pi}}
\cos\left( \kappa r-\frac{l\pi}{2}+\delta_l(\kappa)\right)
\end{equation}
at $r \to\infty$.

Again assuming that $e^{-2S_l}<<1$ one finds that in the region $r>r_3$
\begin{equation}\label{eq:}
\tilde{w}_l(\kappa;r)=- \left(\frac{2}{\pi}\right)^{\frac{1}{2}}\sqrt{\frac{\kappa}{k_l(r)}}
\cos\left(\int_{r_3}^r k_l(r')dr'+\frac{\pi}{4}\right),
\end{equation}
while under the Coulomb barrier
\begin{eqnarray}
\tilde{w}_l(\kappa;r)= -\left(\frac{2}{\pi}\right)^{\frac{1}{2}}\sqrt{\frac{\kappa}{|k_l(r)|}}
\exp\left( \int_{r}^{r_3}|k_l(r')|dr'\right)
\end{eqnarray}
 and inside the nuclear potential well at  $r_1<r<r_2$
\begin{equation}\label{eq:z7}
\tilde{w}_l(\kappa;r)=-\left(\frac{2}{\pi}\right)^{\frac{1}{2}}\sqrt{\frac{\kappa}{k_l(r)}}e^{S_l}
\cos\left(\int_{r}^{r_2} k_l(r')dr'-\frac{\pi}{4}\right).
\end{equation}

Besides, let us introduce the complex functions
\begin{equation}\label{eq:608}
d_l^{(\pm)}(\kappa;r)=w_l(\kappa; r)\pm  i\tilde{w}_l(\kappa;r),
\end{equation}
which are described at  $r\to \infty$  by the running waves:
\begin{equation}\label{eq:605}
d_l^{(\pm)}(\kappa;r) \approx \mp i  \sqrt{\frac{2}{\pi}}
\exp\left\{\pm i\left[\kappa r-\frac{l\pi}{2}+ \delta_l(\kappa) \right ]\right\}.
\end{equation}

 The Eq.~(\ref{eq:t4}) is invariant with respect to reflection of $\kappa$ to $-\kappa$. Therefore its solutions $w_l(\kappa)$ and $\tilde{w}_l(\kappa)$ can only
 change  the sign at such a transformation. Then in correspondence with their boundary conditions  (\ref{eq:t6})
 and (\ref{eq:z6})
 one gets the following
 symmetry conditions:
\begin{eqnarray}\label{eq:606}
w_l(-\kappa;r)=(-1)^{l+1}w_l(\kappa;r),\nonumber \\
\tilde{w}_l(-\kappa;r)=(-1)^{l}\tilde{w}_l(\kappa;r).
\end{eqnarray}
Substitution of (\ref{eq:606}) into (\ref{eq:608}) gives another useful relation
\begin{equation}\label{eq:611}
d_l^{(+)}(-\kappa;r)=(-1)^l d_l^{(-)}(\kappa;r).
\end{equation}

\section{Evolution of the wave packet, which describes alpha-decay}
Let the initial state $\Psi_a(0)=\varphi_a$ of the parent nucleus be formed at $t=0$.
Time-evolution of this wave function  at $t\geq 0$ is governed by the
equation \cite{Goldberger}
\begin{equation}\label{eq:14}
 \Psi_{a}(t)=-\frac{1}{2\pi
i}\int_{-\infty}^{\infty}d\varepsilon e^{-i\varepsilon
t/\hbar}
{\cal G}^{+}(\varepsilon )\Psi_a(0),
\end{equation}
 where  the retarded
Green's operator
 \begin{equation}\label{eq:15}
{\cal G}^{+}(\varepsilon)=(\varepsilon+i\epsilon -H)^{-1},\qquad
\epsilon \to +0.
\end{equation}

The wave function $\Psi_a(t)$ can be expanded in terms of the functions $\varphi_a$ and $\varphi_b^{+}$:
\begin{equation}\label{eq:wff}
\Psi_a(t)=  c_{a}(t)\varphi_a  +\sum_b c_{b}(t)\varphi_b^+,
\end{equation}
where the sum over $b$ denotes the integral over the wave vector ${\boldsymbol \kappa}$   as well as the sum over  quantum numbers of the daughter nucleus $I_d
M_d$.
The expansion coefficients
 are defined by
\begin{equation}\label{eq:111}
c_{a(b)}(t)=-\frac{1}{2\pi
i}\int_{-\infty}^{\infty}d\varepsilon e^{-i\varepsilon
t/\hbar}\langle \varphi^+_{a(b)}|{\cal{G}}^{+}(\varepsilon )|\varphi_a\rangle,
\end{equation}
where the Green matrix is determined by relationships \cite{Goldberger}
\begin{equation}\label{eq:G1}
{\cal G}^{+}_{aa}(\varepsilon)=\frac{1}{\varepsilon-\varepsilon_a-{\cal R}^+_{aa}(\varepsilon)}
\end{equation}
and
\begin{equation}\label{eq:G2}
{\cal G}^{+}_{ba}(\varepsilon)=\frac{{\cal R}^+_{ba}(\varepsilon)}{(\varepsilon+i\epsilon-\varepsilon_b)(\varepsilon-{\cal R}^+_{aa}(\varepsilon))}.
\end{equation}
Here    ${\cal R}^+_{ba}(\varepsilon)={\cal R}_{ba}(\varepsilon+i\epsilon)$ is the  matrix of
the level shift operator, satisfying the integral equation \cite{Goldberger}
\begin{equation}\label{eq:rop}
{\cal R}(\varepsilon)=V+V\frac{1-{\cal Q}_a}{\varepsilon-H_0}{\cal R}(\varepsilon)
\end{equation}
with the  projection operator
\begin{equation}\label{eq:lamda}
{\cal Q}_a=|a\rangle\langle a|
\end{equation}
on  the initial state $|a\rangle$.
Solution of Eq.(\ref{eq:rop}) can be expanded in powers of $V$:
\begin{equation}\label{eq:RM}
 {\cal R}(\varepsilon)=V+V\frac{1-\Lambda_a}{\varepsilon-H_0}V+\cdots.
\end{equation}

The     complex number  ${\cal R}_{aa}^+(\varepsilon_a)$ is usually written down as \cite{Goldberger}
 \begin{equation}\label{}
 {\cal R}^+_{aa}(\varepsilon_a)=D(\varepsilon_a)-i\frac{\Gamma}{2},
 \end{equation}
 where  $D(\varepsilon_a)$ and $\Gamma$ mean the   shift and width of the decaying parent level  (below for brevity a small level shift $D(\varepsilon_a)$   will
 be omitted).
 The total width $\Gamma$ is a sum of all the partial widths:
\begin{equation}\label{eq:Gt}
 \Gamma=\sum_{b}\Gamma_{b}.
\end{equation}
 The partial $\alpha$-decay width reads
\begin{equation}\label{eq:G5}
\Gamma_{b}=2\pi \sum_{M_d}\int d\Omega_{\hat{\boldsymbol \kappa}}|{\cal R}^+_{ba}(\varepsilon_a)|^2\varrho(\varepsilon_b),
\end{equation}
where  the density of final states
\begin{equation}\label{}
\varrho(\varepsilon_b)=\kappa_{b}\mu/\hbar^2
\end{equation}
 depends on   the wave number
\begin{equation}\label{}
\kappa_{b}=\sqrt{2\mu E_{b}}/\hbar.
\end{equation}

The width (\ref{eq:G5}) is proportional to to the squared amplitude $C_l^2$ of the wave $\varphi_b^+$. Therefore according to Eqs.~(\ref{eq:amp}) and
(\ref{eq:res1}) $\Gamma_b\sim e^{-2S_l}$ far from the Bohr-Sommerfeld condition and $\Gamma_b\sim e^{2S_l}$ in the case of resonance as $\Delta\alpha_l=0$. The
latter corresponds to immediate decay of the nucleus and contradicts to all the experimental data.

The Green's functions (\ref{eq:G1}), (\ref{eq:G2})  have a pole in the point $\varepsilon_0={\cal R}^+_{aa}(\varepsilon_a)$ on the second sheet of the complex
Riemann  $\varepsilon$ surface.
Moreover, ${\cal G}^{+}_{ba}(\varepsilon)$ has a pole at $\varepsilon=\varepsilon_b-i\epsilon$.
Inserting the Green's functions
into (\ref{eq:111}) and performing the contour integration (for details see Ref.~\cite{Goldberger}) one arrives at
\begin{eqnarray}\label{eq:p2}
\Psi_a(t)=\varphi_a e^{-i\varepsilon_a t/\hbar-\Gamma t/2\hbar}
 +
\sum_b \frac{{\cal R}^+_{ba}(\varepsilon_a)\varphi_b^+}{\varepsilon_b-\varepsilon_a+i\frac{\Gamma}{2}}
\left[e^{-i\varepsilon_b t/\hbar}-e^{-i\varepsilon_a t-\Gamma t/2\hbar}   \right].
\end{eqnarray}

From here it immediately follows that the probability of finding the parent nucleus in the initial state
 is governed by the exponential decay law \cite{Goldberger}:
\begin{equation}\label{eq:Pa}
P_a(t)=e^{-\Gamma t/\hbar}.
\end{equation}
The probability of finding the clusters at the moment $t$ with  energy in the interval  $(E,\;E+\Delta E)$  can be written
as
\begin{equation}\label{}
\Delta P_d(E, t)=W_d(E, t) \Delta E.
\end{equation}
From Eq.(\ref{eq:p2}) one has the  probability density at $t \to \infty$:
\begin{equation}\label{eq:pe}
W_d(E, \infty)=\frac{1}{\pi}\frac{\Gamma_{d}/2}{(E-E_d)^2+\left(\Gamma/2\right)^2}.
\end{equation}
The corresponding branching ratio will be
\begin{equation}\label{}
\int_{0}^{\infty}W_d(E,\infty)dE=\frac{\Gamma_d}{\Gamma}.
\end{equation}

The level shift operator ${\cal R}(\varepsilon)$ is invariant with respect to rotation. Therefore ${\cal R}(\varepsilon)$ only
 couples   states with the same total angular momentum $I=I_p$ and its projection $M=M_p$. With (\ref{eq:t801})
the off-diagonal matrix elements ${\cal R}_{ba}$ can be represented in the form
\begin{eqnarray}\label{eq:R3}
{\cal R}^+_{ba}(\varepsilon_a)=\kappa^{-1}
\sum_l  \mathfrak{Y}_{I_p}^{M_p}(lI_dM_d;\hat{\boldsymbol \kappa})  \mathscr{I}_{lI_d}(\kappa),
\end{eqnarray}
where $\mathscr{I}_{lI_d}(\kappa)$ denotes the integral
\begin{eqnarray}\label{eq:mi}
\mathscr{I}_{lI_d}(\kappa)=\int d{\bf r}'
\frac{w_l(\kappa;r')}{r'} \int d\xi
\mathscr{Y}_{I_plI_d}^{M_p *}(\xi,\hat{\bf r}')
{\cal R}^+(\varepsilon_a)g_{I_pM_p}(\xi, r').
\end{eqnarray}

Next,  substituting   (\ref{eq:R3})  into (\ref{eq:G5}) one obtains the decay width $\Gamma_{b}$:
\begin{equation}\label{eq:l1}
\Gamma_{b}=2 \pi\sum_{l}|\mathscr{I}_{lI_d}(\kappa)|^2\varrho(\varepsilon_b).
\end{equation}
Here $l$ runs the values from $|I_d-I_p|$ to $I_d+I_p$.
The sum reduces to single term  if $I_p$ or $I_d$  equals  zero.

The relative motion of the clusters in the decay channel $d=I_dM_d$ is described by   the wave function
\begin{equation}\label{}
\Phi_d({\bf r}, t)=\left< g_d g_{\alpha}|\Psi_a(t) \right>e^{i(\varepsilon_b-E)t/\hbar}.
\end{equation}
With the aid of the Eq.~(\ref{eq:p2}) one has
\begin{eqnarray}\label{eq:psi5}
\Phi_{d}({\bf r},t)=\frac{2\mu}{\hbar^2}\int d{{\boldsymbol \kappa}}
\frac{{\cal R}^+_{ba}(\varepsilon_a)\psi^+_{{\boldsymbol \kappa}}({\bf r})}{\kappa^2-\kappa^2_{d}+i\Gamma\mu/\hbar^2}\times
\end{eqnarray}
$$ \times
\left[e^{-iEt/\hbar}-e^{-iE_d t/\hbar-\Gamma t/2\hbar}   \right].
$$

Inserting here the  expressions (\ref{eq:t3}), (\ref{eq:R3}) we first calculate the integral over the spherical angles $\hat{\boldsymbol \kappa}$:
\begin{eqnarray}\label{eq:R1}
\int d\Omega_{{\boldsymbol \kappa}}
{\cal R}^+_{ba}(\varepsilon_a)\psi^+_{{\boldsymbol \kappa}}({\bf r})=
=\sum_{lm}(lI_dm M_d|I_pM_p)\mathscr{I}_{lI_d}(\kappa)\frac{w_l(\kappa;r)}{\kappa^2 r}Y_{lm}(\hat{\bf r}).
\end{eqnarray}
For calculation of the remaining  integral over the radial variable $\kappa$ it is convenient to extend the integration to the whole  region $-\infty<\kappa<
\infty.$
For this aim the denominator of (\ref{eq:psi5}) is rewritten as
\begin{equation}\label{eq:den}
\frac{1}{\kappa^2-\kappa^2_{d}+i\Gamma\mu/\hbar^2} =
\frac{1}{2\kappa_{0}}\left[\frac{1}{\kappa-\kappa_{0}}-
\frac{1}{\kappa+\kappa_{0}}  \right],
\end{equation}
where
\begin{equation}\label{eq:kap}
\kappa_0\approx \kappa_d -i\frac{\Gamma}{2\hbar v_d},
\end{equation}
$v_d=\hbar\kappa_d/\mu$ is the mean velocity of the relative motion of clusters.

Furthermore, it should be taken into account that $\mathscr{I}_{lI_d}(\kappa)w_l(\kappa;r)$ is an even function of $\kappa$
and
\begin{equation}\label{}
w_l(\kappa;r)=\frac{d_l^{(+)}(\kappa;r)+d_l^{(-)}(\kappa;r)}{2},
\end{equation}
Then the  expression (\ref{eq:psi5}) for the wave function  transforms to
\begin{eqnarray}\label{eq:psi7}
\Phi_{I_dM_d}({\bf r},t)=\frac{\phi_{I_dM_d}({\bf r},t)}{r}
\end{eqnarray}
with
\begin{eqnarray}\label{eq:psi8}
\phi_{I_dM_d}({\bf r},t)=-\pi i \frac{\mu}{\hbar^2\kappa_d}\sum_{lm}(lI_d m M_d|I_pM_p)\nonumber \\ \times
\left[{\cal F}^+_l(r,t)+ {\cal F}^-_l(r,t)   \right]Y_{lm}(\hat{\bf r}),
\end{eqnarray}
and
\begin{eqnarray}\label{eq:psi8}
{\cal F}^{\pm}_l(r,t)=\frac{i}{2\pi}\int_{-\infty}^{\infty}\frac{d_l^{\pm}(\kappa;r)\mathscr{I}_{lI_d}(\kappa)}
{\kappa-\kappa_{0}}
\left[e^{-iE t/\hbar}- e^{-iE_d t/\hbar-\Gamma t/2\hbar}\right]d\kappa.
\end{eqnarray}
Here the integration is concentrated mainly around $\kappa_d$ in very narrow interval $\Delta\kappa_d =\kappa-\kappa_d$ of the order of $\Gamma/\hbar v_d$.
 Therefore it is possible to use the approximate equality
\begin{equation}\label{eq:ap}
e^{-iE t/\hbar}\approx e^{-iE_d t/\hbar}e^{-iv_d t\Delta\kappa_d}.
\end{equation}

We shall consider the wave function of the $\alpha$-decay only outside the nucleus, where the smooth  function $\mathscr{I}_{lI_d}(\kappa)$
can be replaced by  $\mathscr{I}_{lI_d}(\kappa_d)$.
Then the function ${\cal F}^{\pm}_l(r,t)$ transforms to
\begin{equation}\label{}
{\cal F}^{\pm}_l(r,t)=\mathscr{I}_{lI_d}(\kappa_d) I^{\pm}_l(r,t)e^{-iE_d t/\hbar},
\end{equation}
where
\begin{eqnarray}\label{eq:psi81}
I^{\pm}_l(r,t)=\frac{i}{2\pi}\int_{-\infty}^{\infty}\frac{d_l^{\pm}(\kappa;r)}
{\kappa-\kappa_{0}}
\left[e^{-iv_d t(\kappa-\kappa_d)}-e^{-\Gamma t/2\hbar}\right]d\kappa.
\end{eqnarray}

The integrand in (\ref{eq:psi81}) has a simple pole  $\kappa_0=\kappa_{d}-i\Gamma/2\hbar v_d$ in the lower
part of the complex $\kappa$ plane. If $v_{d}t<r$   the integration contour in
$I^{+}_l(r,t)$ is supplemented by a semicircle $C$ of the radius $R \to \infty$ in the upper half-plane $\kappa =\kappa'+i\kappa''$. If $v_{d}t>r$
it is done in the lower half-plane. The integration along these  semicircles $C$, where $d_l^{\pm}(\kappa,R) \sim e^{\pm i\kappa R}$, gives zero.
Then the integral $I^+_l$ is easily calculated by means of the residue theory.
 As to the function  $I^{-}_l(r,t)$, it appears to be proportional to the difference of two exponents  $e^{-\Gamma t/2\hbar}$
at any moment $t$.  Hence $I^{-}_l(r,t)=0.$

 Finally the wave function outside the nucleus takes the form
\begin{eqnarray}\label{eq:222}
\phi_{I_dM_d}({\bf r},t)=-i\sqrt{\frac{\pi}{2}}\sum_{lm}{\cal A}^{(lm)}_{I_dM_d}
d_l^{(+)}(\kappa_{0};r)
Y_{lm}(\hat{\bf r})\nonumber \\ \times
 \exp\left[-iE_d t/\hbar-\Gamma t/2\hbar\right]\Theta(t-r/v_{d}),
\end{eqnarray}
where the amplitude
\begin{equation}\label{}
{\cal A}^{(lm)}_{I_dM_d}=-\sqrt{2\pi}\frac{\mu}{\hbar^2\kappa_d}(lI_dmM_d|I_pM_p)\mathscr{I}_{lI_d}(\kappa_{d}),
\end{equation}
and the Heaviside step function
\begin{equation}\label{}
\Theta(x)=\left\{\begin{array}{lll} 1,&\;&\;x>0,\\
0,&\;&\;x<0.  \end{array}\right.
\end{equation}

Notice that $d_l(\kappa_0;r)$ in (\ref{eq:222}) depends on the
complex number $\kappa_0$, defined by Eq.~(\ref{eq:kap}).
Therefore in the asymptotic region, where  $d_l^{+}(\kappa_{0};r)$
is presented by Eq.~(\ref{eq:605}), the wave function takes the
form
\begin{eqnarray}\label{eq:223}
\phi_{I_dM_d}({\bf r},t)=\sum_{lm}{\cal A}^{(lm)}_{I_dM_d}
e^{i(\kappa_d r-l\pi/2+\delta_l)}
Y_{lm}(\hat{\bf r})\nonumber\\ \times
 \exp\left[-iE_d t/\hbar-\Gamma (t-t_0)/2\hbar\right]\Theta(t-t_0),
\end{eqnarray}
where $t_0=r/v_d$ is the arrival time of $\alpha$-particles in the point $r$.

We see that the emitted $\alpha$-particles are described by a spherically  diverging wave, which propagates with the wave vector $\kappa_d$ and has a sharp wave
front, defined by the radial coordinate $r_f=v_d t$.
The intensity of this wave ${\cal I}_d(r,t)=|\phi_{I_dM_d}(r,t)|^2$  in units of $|\phi_{I_dM_d}(r_f,t)|^2$ can be written as
 \begin{figure}[h]
\vspace{0cm}
\centerline{\includegraphics[height= 10cm, width= 10cm]{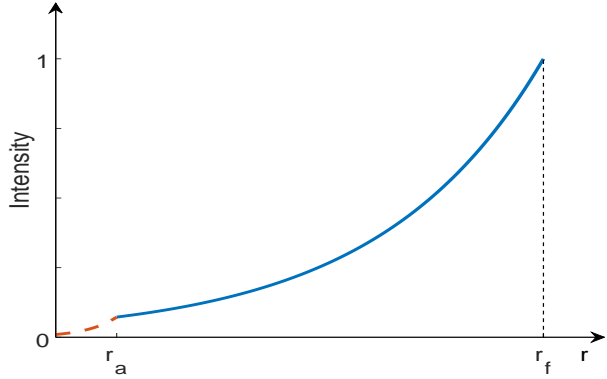}}
\vspace{-2cm}
\caption{\label{fig2} Intensity dependence of the wave function, which describes relative motion of the decay fragments, on the relative coordinate $r$ at
$t=2\tau_n$.  The coordinate $r_f=v_d t$ determines position  of the wave front, $r_a$  the atom radius.}
\end{figure}
\begin{equation}\label{}
{\cal I}_d(r,t)=
\exp\left[-\frac{t}{\tau_n}\left(1-\frac{r}{r_f}\right)  \right],
\end{equation}
where $\tau_n=\hbar/\Gamma$ is the nuclear lifetime.
Dependence of the intensity ${\cal I}_d(r,t)$ on $r$ at fixed moment $t=2\tau_n$ is shown in Fig.2.

The detection  probability of the $\alpha$-particle at the moment $t$ in the $d$th channel
reads
\begin{equation}\label{eq:224}
P_d(t)=\sum_{M_d}\int|\Phi_{I_dM_d}({\bf r},t)|^2d{\bf r}.
\end{equation}
For large enough times, when $t>>r_a/v_{d}$,  the main contribution into this integral is from the  region outside the atom
of the radius $r_a$.
Then substituting the asymptotic expression (\ref{eq:223}) into (\ref{eq:224}) one immediately gets
\begin{equation}\label{eq:p1}
P_{d}(t)=\frac{\Gamma_{d}}{\Gamma}\left(1-e^{-\Gamma t/\hbar}\right).
\end{equation}
According to Eq.~(\ref{eq:Gt}) the sum of  the decay probabilities into all possible channels and the survival probability
(\ref{eq:Pa}),   at any moment $t$ equals unity, that confirms the correctness of above calculations.

\section{Moshinsky's function}
In the asymptotic region one can avoid  the approximation (\ref{eq:ap}) and express  the wave function in terms of the  Moshinsky function $M(r,\kappa_0,t)$ [15-19]:
\begin{eqnarray}\label{eq:225}
\phi_{I_dM_d}({\bf r},t)=
   \sum_{lm}{\cal A}^{(lm)}_{I_dM_d} M(r,\kappa_0,t)
      e^{-il\pi/2+i\delta_l(\kappa)}
Y_{lm}(\hat{\bf r}).
\end{eqnarray}
The Moshinsky function is given by
\begin{eqnarray}\label{eq:Mosh}
M(r,\kappa_0,t)=\frac{i}{2\pi}\int_{-\infty}^{\infty}\frac{e^{-i\hbar\kappa^2t/2\mu}e^{i\kappa r}}{\kappa-\kappa_0}d\kappa \nonumber\\
=\frac{1}{2}e^{-i\hbar\kappa_0^2 t/2\mu}e^{i\kappa_0 r}\mbox{erfc}(y),\qquad  \mbox{Im}\;\kappa_0<0,
\end{eqnarray}
where  $\mbox{erfc}(y)$ is the complementary error function:
\begin{equation}\label{eq:er}
 \mbox{erfc}(y)=1-  \mbox{erf}(y),
\end{equation}
expressed in terms of the error function
\begin{equation}\label{eq:er}
\mbox{erf}(y)= \frac{2}{\sqrt{\pi}}\int_{0}^{y}e^{-u^2}du,
\end{equation}
depending on
\begin{equation}\label{eq:def}
y=e^{-i\pi/4}x, \qquad x=\left(\frac{\mu}{2\hbar t}\right)^{1/2}\left(r-v_0 t \right), \qquad v_0=\hbar\kappa_0/\mu.
\end{equation}

Bearing in mind Eq.~(\ref{eq:kap}),  one can reduce the function (\ref{eq:Mosh})  to
\begin{equation}\label{eq:M9}
M(r,\kappa_0,t)=e^{i\kappa_d r-iE_d t/\hbar}e^{-\Gamma(t-r/v_d)/2\hbar}\frac{1}{2}\mbox{erfc}(y),
\end{equation}
where $v_0$ is replaced by $v_d$ since  $\Gamma<<E_d$.
At last, making  substitution (\ref{eq:M9}) into (\ref{eq:225}) one arrives at the same expression (\ref{eq:222}) for the $\alpha$-wave function, but
with  $(1/2)\mbox{erfc}(y)$ instead of $\Theta(t-r/v_d)$.

At the wave front $r=r_f=v_dt$ the complementary error function $\mbox{erfc}(0)$ by definition equals unity.
For analysis  of its behavior at large values of $|y|$ let us
make the substitution $u=e^{-i\pi/4}\varsigma$, giving
 \begin{equation}\label{}
 \mbox{erf}(y)=\sqrt{ \frac{2}{\pi}}(1-i)\left(\int_{0}^{x}\cos \varsigma^2 d\varsigma+i \int_{0}^{x}\sin \varsigma^2 d\varsigma   \right).
 \end{equation}
 At $|x|\to \infty$ these integrals reduce to the table integrals \cite{RG}
 \begin{equation}\label{}
 \int_{0}^{\infty}\cos \varsigma^2 d\varsigma= \int_{0}^{\infty}\sin \varsigma^2 d\varsigma=\frac{1}{2}\sqrt{\frac{\pi}{2}},
 \end{equation}
leading us  to the conclusion that
\begin{equation}\label{}
\lim_{x \to \infty}\mbox{erfc}(y)=0, \qquad \lim_{x \to -\infty}\mbox{erfc}(y)=2.
\end{equation}

Thus, far from the point $y=0$ the  $(1/2)|\mbox{erfc}(y)|$ coincides with the Heaviside step function. Numerical calculations of
 $(1/2)|\mbox{erfc}(y)|$ are displayed in Fig.3. They demonstrate that   the wave front is considerably distorted only in the interval $|x|\sim 10$.
  \begin{figure}[h]
\vspace{0cm}
\centerline{\includegraphics[height= 7cm, width= 8cm]{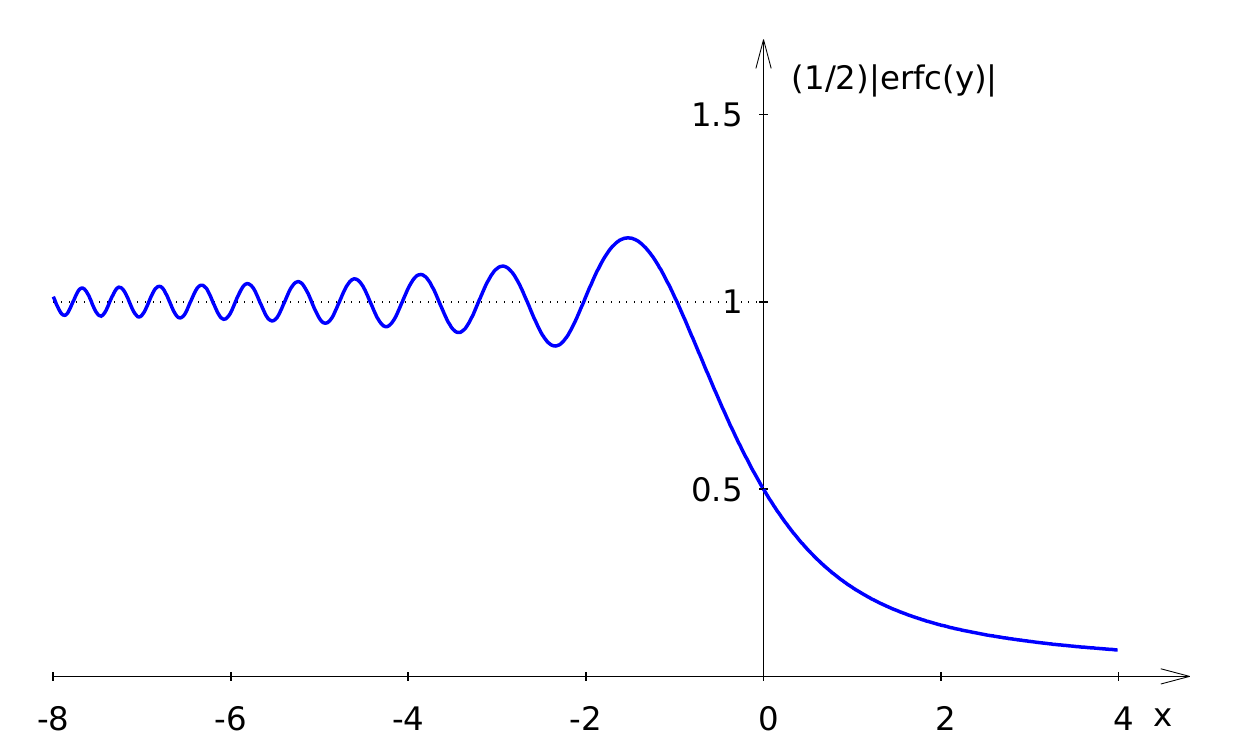}}
\vspace{0cm}
\caption{\label{fig3} Dependence of the function $(1/2)|\mbox{erfc}(y)|$, where $y=e^{-i\pi/4}x$, on the variable $x$.}
\end{figure}
 It corresponds
to the time interval $\Delta  t=t-t_0$ of the order of
\begin{equation}\label{}
|\Delta t| \sim \frac{10}{v_d}\sqrt{\frac{2r}{\kappa_d}},
\end{equation}
where  $t_0=r/v_d$ designates  the mean arrival moment of the
$\alpha$-wave packet  to the detector, located  in the point $r$.
For the distance from the target to detector $\sim 1$ m and the
energy $E_d \sim 5$ MeV we get the arrival time  $t_0 \sim
10^{-6}$ s and the time interval $|\Delta t|\sim 10^{-14}$ s.
Unfortunately, such tiny time window makes unreal
 observation of the Moshinsky transient effect in the $\alpha$-decay.

\section{Approximate calculations}
The Eq.~(\ref{eq:l1}) is unpractical for numerical calculations of the decay constant $\lambda_d=\Gamma_d/\hbar$. Therefore let us make some  simplifications,
taking into account that
 it is proportional to the probability
of finding the $\alpha$ particle inside the nucleus. In the case when the emitted $\alpha$-particle carries away single orbital momentum $l$, this probability  is
equal to the integral of $w^2_l(\kappa;r)/(\kappa r)^2$ over the nuclear volume.
Therefore neglecting the exponentially attenuating tail of $w_l(\kappa;r)$ under the centrifugal barrier, one can represent the squared modulus of the matrix
element (\ref{eq:mi}) as
\begin{equation}\label{}
|\mathscr{I}_l|^2=\mathscr{N}^2\int_{r_1}^{r_2}\frac{w^2_l(\kappa_l;r)}{\kappa_l^2}dr,
\end{equation}
where the coefficient $\mathscr{N}$  has dimensionality of energy.
Then the decay constant  transforms to
\begin{eqnarray}\label{eq:fin}
\lambda_l=\mathscr{N}^2\frac{4\mu}{\hbar^3} \frac{e^{-2S_l}}{\sin^2\alpha_l} \int_{r_1}^{r_2}
\frac{dr}{k_l(r)} \cos^2\left(\int_{r_1}^r k_l(r')dr' -\frac{\pi}{4} \right).
\end{eqnarray}
As usually (see, e.g., Ref.~\cite{Gurvitz}) the quickly oscillating squared cosine is replaced by 1/2 giving
\begin{equation}\label{eq:2AM}
T_l=\frac{2\mu}{\hbar}\int_{r_1}^{r_2}\frac{dr}{k_l(r)},
\end{equation}
which may be interpreted as a quasi-classical period of the $\alpha$-particle  oscillations inside the nucleus between the turning points $r_1$ and $r_2$.
The corresponding knocking frequency is $\nu_l=1/T_l$.

Then (\ref{eq:fin}) is reduced to standard expression
\begin{equation}\label{eq:lam}
\lambda_l=p_l\nu_l e^{-2S_l },
\end{equation}
 where   the factor $p_l$ is given by
 \begin{equation}\label{eq:p}
p_l=\frac{8\pi^2}{\sin^{2}\alpha_l}\left(\frac{\mathscr{N}}{\hbar\omega }\right)^2,
\end{equation}
 $\omega=2\pi \nu$ and $\hbar\omega $ can be interpreted as the phonon energy  of the $\alpha$-particle vibrations.

\section{Conclusion}
So a straightforward solution of the time-dependent Schr\"{o}dinger equation is reported for the $\alpha$-decay.  It is based on the idea that the parent nucleus,
formed at $t=0$ in any nuclear process, occurs in the bound state $\varphi_a$, described by the shell model, treating  all the nucleons as an ideal gas. Every
occupied nuclear level has negative energy, lying lower than the continuous spectrum. Therefore the unperturbed  wave function $\varphi_a$ corresponds to really bound state.
 Such a function is orthogonal to the scattering functions $\varphi_b^+$ of the continuous spectrum, which  describe the $\alpha$-particle and daughter
 nucleus. And only the residual interaction $V'$ gives rise to the exponential  decay of the state $\varphi_a$, coupling it with the states  $\varphi_b^+$.

 The value of the isolated energy level  $\varepsilon_a$ of the initial state is dictated by the character of the nuclear forces,  but not
 by our desire to fulfill the Bohr-Sommerfeld quantization rule (\ref{eq:Som}). Remind that this requirement determines  the resonance levels $E_r$ in the
 scattering of particles by any potential well \cite{Sitenko}. At these energies the wave function of the incident particle inside the well
 reaches maximum, while the
  scattering cross section  attributes a bump with the width $\Gamma_r$. This fact is well illustrated for the scattering of $\alpha$-particles by numerical
  calculations \cite{Fedor}. Furthermore, if at $t=0$ there is an $\alpha$-wave packet inside the nucleus, which is spread in energy interval $\Delta E >>
  \Gamma_r$ and concentrated at the resonance energy $E_r$, then later it exponentially decays with the lifetime $\tau=\hbar/\Gamma_r$ \cite{Sitenko}.
It is curious, that the decay constant $\lambda=1/\tau$ for such a quasistationary level, derived in Eq.~(\ref{eq:res}), is easily transformed to the Gamov's
formula $\lambda=\nu e^{-2S}$ with the knocking frequency $\nu$ being reciprocal to the period of vibrations $T$, standing in Eq.~(\ref{eq:2AM}). Such kind of
resonances is provided by the interference of the waves reflected by the edges of the potential well \cite{Bohm}.

At the same time, there should be resonances in the $\alpha$-particle  scattering, caused by the reconstruction of internal nuclear  structure. Namely, the
resonances arising when the $\alpha$-particle is captured into the compound nuclear state, and its energy  is shared among all the nucleons of the compound
nucleus. They are reproduced by the famous  Breit-Wigner's formula. However, this resonant $\alpha$-scattering cross section is proportional to $e^{-2S}$, so that
it is impossible to observe it at low temperatures, when the barrier penetrability is too small.
In the reciprocal process of the $\alpha$-decay such levels of the parent (compound) nucleus manifest themselves  as long-living states.

According to the derived Eqs.~\ref{eq:mi}), (\ref{eq:l1}) the $\alpha$-decay width $\Gamma$ is  proportional to the squared amplitude of the wave function
$C_l^2$.
Moreover,  Eqs.~(\ref{eq:amp}), (\ref{eq:cs}) show that $C_l^2\sim e^{-2S_l}$ far from the "interference" resonance, while $C_l^2\sim  e^{2S_l}$ if the  condition
(\ref{eq:Som}) for this resonance is exactly fulfilled. Hence, only far from the "interference" quasistationary level we come to the correct result $\lambda \sim e^{-2S}$, while in the
resonance the decay becomes practically instantaneous. In the case of  decay deeply under the barrier the probability for the energy $\varepsilon_a$ of the parent
nucleus to fit the narrow 'interference" resonance window $\Gamma \sim  e^{-2S_l}$ is too tiny to be realized in nature.

In standard approach to the problem  the stationary Schr\"{o}dinger equation is solved, resulting in the complex energy and exponentially diverging wave function,
which is spread over the whole space. Its amplitude at the nuclear surface is of the order of unity. On the contrary, the partial   scattering wave functions
$w_l(\kappa,R)$, having the amplitude $C_l \sim e^{-S_l}$ far from the Bohr-Sommerfeld condition, are weak inside the nucleus, then exponentially grow under the
Coulomb barrier, and approach their asymptotic (\ref{eq:t6})  outside the atom $(r>r_a)$. Given by Eq.~(\ref{eq:222}), the complete wave packet  $\Phi(r,t)$    is
formed by superposition of these basis functions. Exponentially growing with $r$,  the $\Phi(r,t)$ is  truncated at the point $r_f=v_d t$, that ensures its proper
normalization. Distortions of  the  $\alpha$-wave front, calculated by means of strict Moshinsky's approach, appear in very narrow
time interval  $\Delta t$ and therefore  can be ignored in the experiment.

\end{document}